\documentclass[aps,groupedaddress,prb,reprint,showpacs]{revtex4-1}
\usepackage{graphicx}
\usepackage{amssymb}
\begin{document}

\title{Electron backscattering from stacking faults in SiC by means of \textit{ab initio} quantum transport calculations}

\author{I. Deretzis}
\email{ioannis.deretzis@imm.cnr.it}
\affiliation{Istituto per la Microelettronica e Microsistemi (CNR-IMM), Z.I. VIII Strada 5, I-95121 Catania, Italy}

\author{M. Camarda}
\affiliation{Istituto per la Microelettronica e Microsistemi (CNR-IMM), Z.I. VIII Strada 5, I-95121 Catania, Italy}

\author{F. La Via}
\affiliation{Istituto per la Microelettronica e Microsistemi (CNR-IMM), Z.I. VIII Strada 5, I-95121 Catania, Italy}

\author{A. La Magna}
\affiliation{Istituto per la Microelettronica e Microsistemi (CNR-IMM), Z.I. VIII Strada 5, I-95121 Catania, Italy}
\date{\today}

\begin{abstract}
We study coherent backscattering phenomena from single and multiple stacking faults (SFs) in 3C- and 4H-SiC within density functional theory quantum transport calculations. We show that SFs give rise to highly dispersive bands within both the valance and conduction bands that can be distinguished for their enhanced density of states at particular wave number subspaces. The consequent localized perturbation potential significantly scatters the propagating electron waves and strongly increases the resistance for $n$-doped systems. We argue that resonant scattering from SFs should be one of the principal degrading mechanisms for device operation in silicon carbide. 
\end{abstract}
\pacs{61.72.Nn,72.10.Fk,73.40.-c}

\maketitle



Silicon carbide is a wide band-gap semiconductor with a formidable number of stable polytypes ($>$ 200) that are distinguished by a unique stacking sequence of SiC bilayers. The origin of this vast polymorphism stems from the small differences in the formation energies between different stacking sequences that outscore all other semiconductors by at least one order of magnitude \cite{Choyke,1983RSPSA.386..241P,1978PSSAR..45..207G,1984PMagA..50..171T}. Likely, the same reason lies underneath the frequent presence of stacking faults (SFs), i.e. irregularities in the stacking sequence, found in this material. Even if the structural lattice deformation induced by  SFs is marginal, their electronic response is that of localized defects; they introduce particular quantum-well-like states in the eigenspectrum, with wave functions that are confined within $\sim$10 \AA{} around the SF \cite{2002PhRvB..65c3203I,2003PhRvB..67o5204L,2011APExp...4b5802C}. Since the early days of SiC device fabrication, there was a clear evidence that such defects were related to a degradation of the device operation, usually giving leakage currents during electrical conduction \cite{2006JAP....99a1101S}. However, such vision overlooks the possibility of SFs  acting as scattering centers during the transport process \cite{1990PhRvB..41.5280S}. From a fundamental point of view, the presence of a localized potential barrier in a material (e.g. from a defect or impurity) can either partially or totally reflect the incident electronic waves, on the basis of its height and energy resonance \cite{2010PhRvB..81h5427D}. Stacking faults can be no exception to this. Rather, their plausible micrometer dimensions can make this aspect crucial. In this case the perturbation induced on the transmitting waves is not localized at an atomic position (like in the case of vacancies, interstitials, etc.), but it is spatially extended throughout the entire width of the dislocation.

 \begin{figure}
	\centering
		\includegraphics[width=0.9\columnwidth]{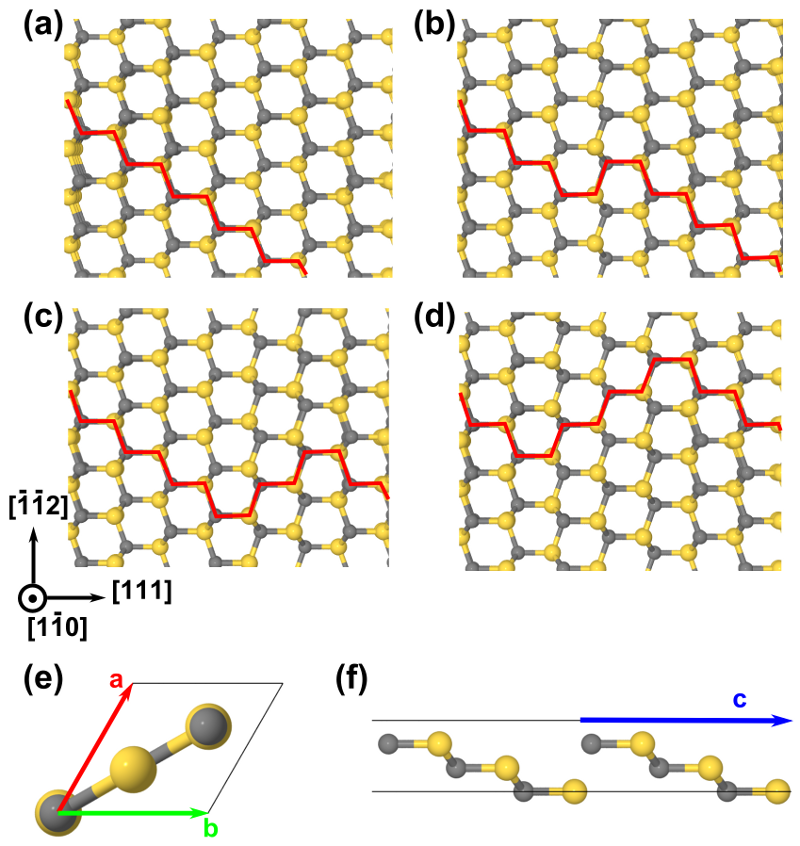}
	\caption{Crystal structure of 3C-SiC in the presence of SFs: (a) ideal, (b) single SF, (c) double SF and (d) triple SF (micro-twin). Transverse (e) and parallel (f) views with respect to the transport direction, showing the lattice vectors of the hexagonal unit cell.}
	\label{fig:sfs}
\end{figure}

In this article we study the coherent backscattering mechanism induced by stacking faults in cubic and hexagonal SiC within a computational approach that treats the electronic structure from first principles (using the density functional theory) and the quantum transport within the nonequilibrium Green's function formalism \cite{Datta}. We show that the effect of this scattering mechanism can be extremely important for the conduction-band transport characteristics, giving rise to total electronic reflections for particular resonances and types of defects.

We commence by simulating three-dimensional 3C-SiC systems with single, double and triple (micro-twin) SFs perpendicular to the [111] crystallographic direction [Fig. \ref{fig:sfs}(a-d)] using the supercell method \cite{2003PhRvB..67o5204L,1985PhRvB..32.7979C}. We construct the supercells using a hexagonal symmetry for the (111) plane while the c-axis is parallel to the [111] direction [Fig. \ref{fig:sfs}(e-f)]. The c-axis also defines the direction of transport, making the incident electron angle $\phi$ with respect to the SF equal to $\phi=90^{\circ}$. We consider a single SF within the supercell, ensuring that the cell length guarantees the absence of SF interactions between successive periodic images. We calculate the electronic structure using the SIESTA code\cite{2002JPCM...14.2745S} along with the Local Density Approximation (LDA) \cite{1981PhRvB..23.5048P} for the electronic correlations, in order to have a clear comparative framework with Ref. \citenum{2002PhRvB..65c3203I,2003PhRvB..67o5204L,1985PhRvB..32.7979C}. The wave functions are written on a localized basis set of double-$\zeta$ valence plus polarization atomic orbitals for both Si and C, while the electronic contribution of the ionic cores is described with norm-conserving Troulier-Martins pseudopotentials \cite{1991PhRvB..43.1993T}.  The minimization of the electron density is achieved by sampling the Brillouin zone with an  $8 \times 8 \times 1$ Monkhorst-Pack grid, whereas a mesh cutoff energy of 450 $Ryd$ is used for real-space integrals. Structural relaxation is obtained with a force criterion of 0.04 eV/\AA{}. Similar calculations are also conducted for silicon.

Upon convergence, the Hamiltonian ($H$) and Overlap ($S$) matrices are extrapolated for 140 nonequivalent $k$-points along the $\Gamma \rightarrow M \rightarrow K \rightarrow \Gamma$ Brillouin-zone path, which is parallel to the SF junction ($k_{\parallel}$  from now on). Quantum transport is calculated for the ballistic regime, starting from the single-particle retarded Green's function matrix 

\begin{equation}
\mathcal{G}^r_{k_{\parallel}}(\epsilon) = [ \epsilon S_{k_{\parallel}} - H_{k_{\parallel}} - \Sigma_{L,k_{\parallel}} - \Sigma_{R,k_{\parallel}} ]^{-1},
\end{equation}

where $\epsilon$ is the energy and $\Sigma_{L,R}$ are self-energies that account for the effect of ideal semi-infinite contacts (obtained from surface Green function techniques \cite{1985JPhF...15..851L}). The transmission probability of an incident Bloch state with energy $\epsilon$ and momentum $k_{\parallel}$ can be thereon computed as:

\begin{equation}
T_{k_{\parallel}}(\epsilon)=Tr \{ \Gamma_{L,k_{\parallel}} \mathcal{G}^r_{k_{\parallel}} \Gamma_{R,k_{\parallel}} [ \mathcal{G}^r_{k_{\parallel}} ]^\dagger \},
\end{equation}

where $\Gamma_{L(R),k_{\parallel}}=i \{ \Sigma_{L(R),k_{\parallel}}- [\Sigma_{L(R),k_{\parallel}}] ^\dagger \}$ are the spectral functions of the contacts. Consequently, the reflection coefficient can be defined as $R = 1 - T$. According to the Landauer-Buttiker theory \cite{Datta}, conductance can be calculated as $G=G_0T$, where $G_0=2e^2/h \approx 77.5 \mu S$ is the conductance quantum. Similarly, from the spectral function of the scattering region $A_{k_{\parallel}} = i \{ \mathcal{G}^r_{k_{\parallel}}- [ \mathcal{G}^r_{k_{\parallel}} ]^{\dagger} \}$ the k-resolved density of states (DOS) can be computed as:

\begin{equation}
DOS_{k_{\parallel}}(\epsilon) = Tr [ A_{k_{\parallel}} S_{k_{\parallel}} /(2 \pi) ]
\label{eq:dos}
\end{equation}

It has to be noted here that the total transmission and DOS in real space can be estimated for each energy by integration over the reciprocal space. In this case, the number of the sampled $k$-points for each Brillouin zone subpath becomes proportional to the dimension of the corresponding real-space direction, being a measure of confinement for the crystal.

 \begin{figure}
	\centering
		\includegraphics[width=\columnwidth]{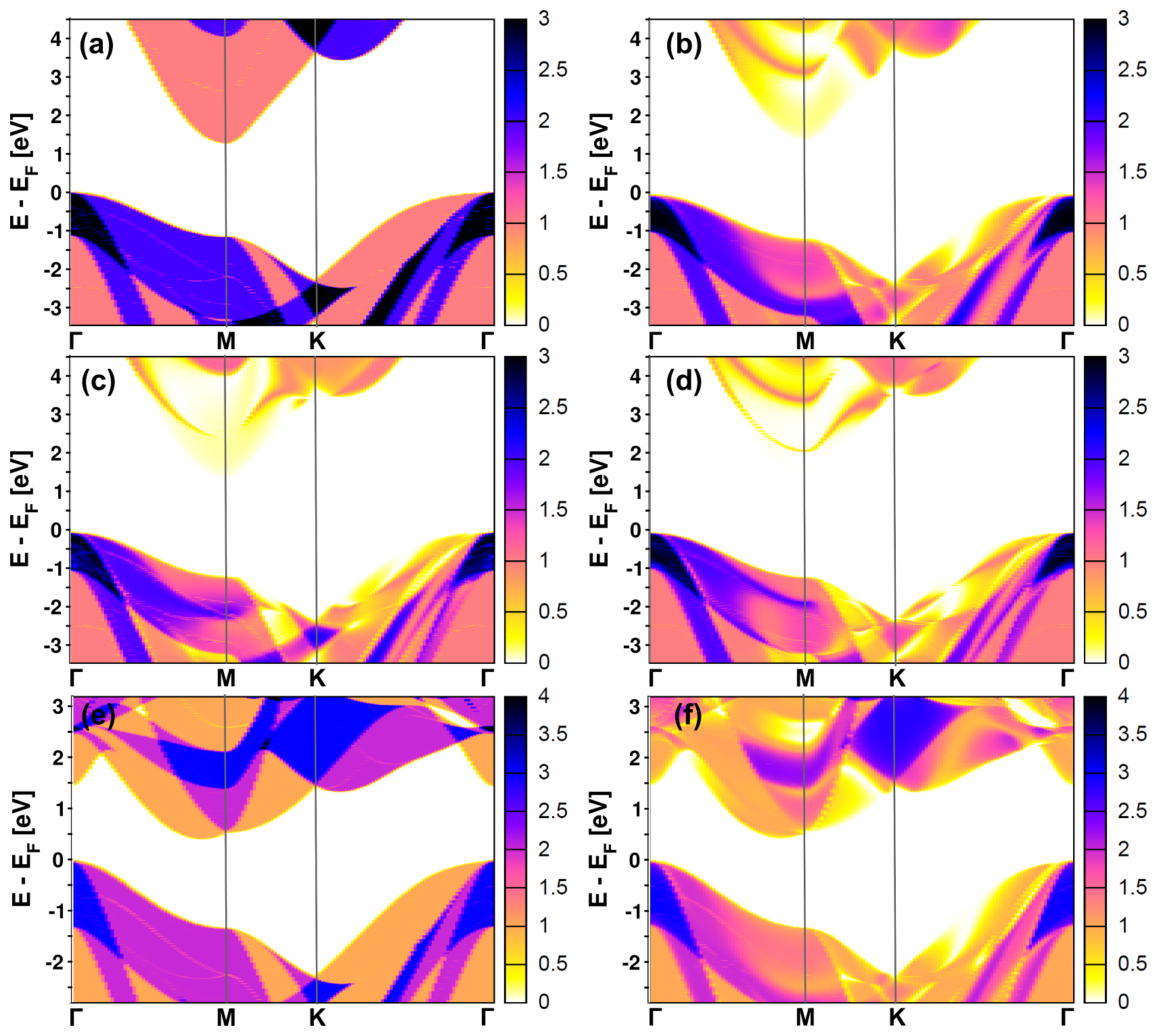}
	\caption{Momentum-space projection of the conductance as a function of energy (in units of $G_0=2e^2/h$) for (a) ideal 3C-SiC, (b) 3C-SiC with a single SF, (c) 3C-SiC with a double SF, (d) 3C-SiC with a micro-twin SF, (e) ideal Si and (f) Si with a singe SF.}
	\label{fig:3ccond}
\end{figure}

\begin{figure}
	\centering
		\includegraphics[width=\columnwidth]{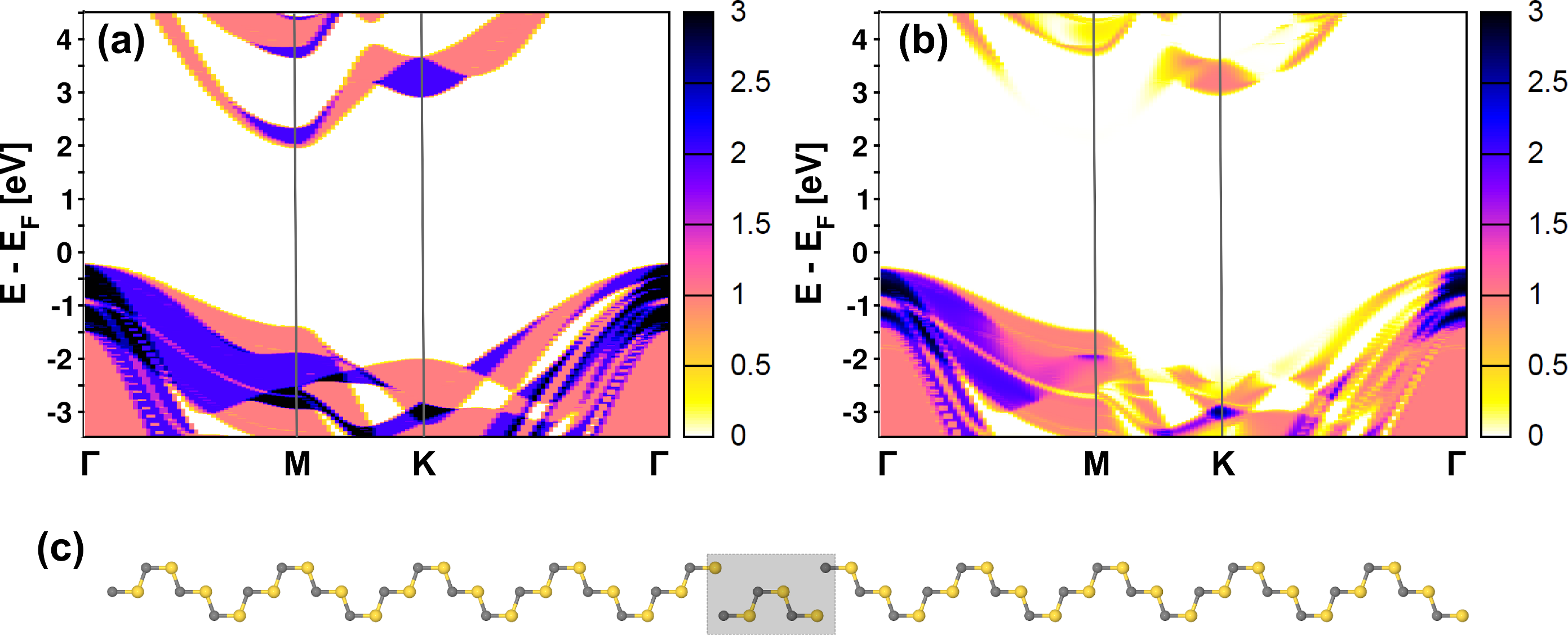}
	\caption{Momentum-space projection of the conductance as a function of energy (in units of $G_0=2e^2/h$) for (a) ideal 4H-SiC, (b) 4H-SiC with a (4,4) SF. (c) Schematic representation of the 4H-SiC supercell, highlighting the (4,4) SF.}
	\label{fig:4hcond}
\end{figure}

We start by considering a momentum-space projection of the main transport attributes of 3C-SiC. The elongated structure of the supercell along the [111] direction gives rise to a quasi-two-dimensional Brillouin zone with the $\Gamma$ point lying at its center, whereas the $M$ and $K$ points are positioned at the center and the angle of the hexagon's side respectively\citep{1985PhRvB..32.7979C}. The $k$-resolved ideal conductance of pure 3C-SiC [Fig. \ref{fig:3ccond}(a)] has some particular characteristics with respect to these high-symmetry points: (i) the valence band maximum is mapped at the $\Gamma$ point whereas  the conduction band minimum is mapped at the $M$ point. Therefore these are the points of interest for low-bias $p$- and $n$-type conduction respectively. (ii) There exists only a single conduction channel per $k$-point for the conduction band up to $\sim$2 eV from its minimum, as can be deduced by the $G_{k_{\parallel}}=G_0$ value of the conductance for this energy range. Contrary, the conductance value at the maximum of the valence band is $3G_0$.  This  feature makes cubic SiC particular within the SiC polytypes, since the breaking of the cubic symmetry results in higher values of the conductance at the minimum of the conduction band (e.g. in the case of 4H-SiC $G_{k_{\parallel}}=2G_0$ near the conduction band minimum [see Fig. \ref{fig:4hcond}(a)]). The presence of a single SF  [Fig. \ref{fig:3ccond}(b)] drastically changes this picture. Band structure calculations already exclude the possibility of intra-gap states from SFs in 3C-SiC, contrary to other hexagonal polytypes \citep{2003PhRvB..67o5204L}.  SF-induced states in this case are introduced inside the valence and conduction bands. Their influence on the system is perturbative and as a result, the transport channels get partially blocked. The local effective potential induced by the single-SF scatters the electron waves mainly around the  $M$ point at the conduction band and around the $K$ point in the valence band. Hence, scattering by the SF has an important effect on the low-bias conductance of $n$-type systems and a marginal influence on the conductance of $p$-type systems. The percentage of reflection is $86\%$ near the conduction band minimum, meaning that the single-SF barrier strongly increases the resistance, albeit not completely suppressing it. This picture quantitatively changes in the case of the double and the triple SFs [Fig. \ref{fig:3ccond}(c-d)]. Here the increase of the barrier height due to the spatial extension of the defect in real space gives rise to an almost complete reflection of the electron waves at the bottom of the conduction band [see table \ref{tab:reflect}], and transport effectively takes place only for higher energies. In this case, this extreme enhancement of the resistance can be directly related to a respective increase of the measured Schottky-barrier height for $n$-type 3C-SiC systems; indeed, even if the difference between the electronic states at the conduction band minimum and the Fermi level is not altered by the SF, electrical measurements are expected to measure the effective height between the Fermi level and the first states within the conduction band that do not totally reflect the electrons. Our calculations indicate that the theoretical limit for this increase could be as high as $\sim 0.8 eV$ with respect to the ideal case for charges approaching the SFs at right angles; however, considering the presence of electron dephasing in real systems, the measured value is expected to be lower. For all the previous cases, transport at the top of the valence band practically remains unaffected, implying that resonant backscattering from SFs should not significantly influence conduction for $p$-type 3C-SiC systems. 

The origin of this strong scattering mechanism in 3C-SiC is inherently related to the geometrical symmetry of the SFs (being perpendicular to the [111] direction) with respect to the directionality of the atomic orbitals that form the wave functions at the minimum of the conduction band (considering a linear wave function expansion on an atomic orbital basis). A direct comparison with the case of Si (ideal and with a single SF [Fig. \ref{fig:3ccond}(e-f)]) is illuminating: a first reading shows that the main difference between the two systems lies in their conduction bands, which are clearly inequivalent. If we now examine the single-orbital contributions in the wave functions for each system \cite{2011PhRvB..84w5426D}, we find that for Si, the conduction band minimum has principal contributions from atomic orbitals that are not polarized towards the [111] direction. Hence, the SF does not break the symmetry of the respective wave function and as a result reflection is almost absent. Contrary, in the case of 3C-SiC, orbitals with a directionality parallel to the c-axis do contribute in the formation of the wave functions at the conduction band minimum. In this case the geometrical symmetry breaking is reflected into an electronic symmetry breaking, which forms the origin of the conductance perturbation. Moreover, for higher energies, silicon has numerous bands throughout the entire Brillouin zone, which all contribute in the total conductance of the system. Based on the polarization criterion, the states introduced by the SF affect only some of them, whereas others remain unaffected. As a result reflection is overall moderate.

\begin{figure}
	\centering
		\includegraphics[width=\columnwidth]{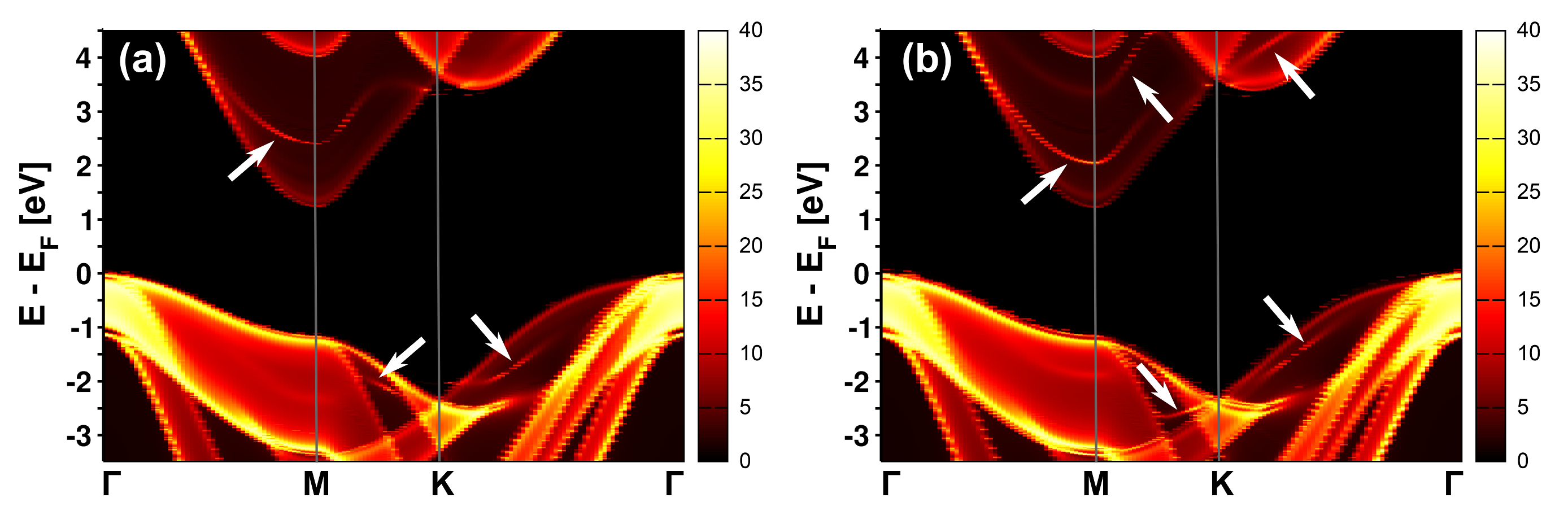}
	\caption{Momentum-space projection of the density of states as a function of energy (in arbitrary units) for (a) 3C-SiC with a double SF and (d) 3C-SiC with a micro-twin SF. Arrows indicate the bands that originate from the SFs.}
	\label{fig:dos}
\end{figure}

A key point that emerges with respect to the transport features presented in the previous paragraphs is the nature of the SF states in 3C-SiC. Previous studies \citep{2002PhRvB..65c3203I,2003PhRvB..67o5204L} have indicated that these states should have resonances at the bottom of the conduction band and excluded any intra-gap contribution. However their degree of localization and mixing with the non-localized bands of the crystal make their exact identification unclear for band structure calculations, and similarly difficult for experimental measurements. In our approach, based on an exact estimation of the density of states in the reciprocal space (see eq. \ref{eq:dos}), we show that SFs give rise to highly dispersive bands within both the conduction and the valence zones. These bands are characterized by an enhanced density of states within particular wave number subspaces, which correspond to the Brillouin-zone regions of enhanced backscattering during conduction [Fig. \ref{fig:dos}(a-b)]. Table \ref{tab:reflect} shows the resonant energies of the various SFs studied in the previous paragraph for 3C-SiC at the $M$ Brillouin-zone point. A common aspect for all SFs is that they generate a state that is very close to the conduction band border with a ``faint" DOS signature. Moving onwards to higher energies, SF bands are characteristic of the SF type and have a higher DOS with respect to the non-defected crystal. 

\begin{table}
	\centering
		\begin{tabular*}{\columnwidth}{@{\extracolsep{\fill}} l c c  }
		\hline 
		\hline 
		System & Reflection coeff. (\%) & $E_M$ (eV) \\
		\hline
	3C-SiC single SF & 86 & 0.1/1.9 	\\
	3C-SiC double SF & 98 & 0.1/1.2 	\\
	3C-SiC triple SF & 100 & 0.2/0.8/2.2 	\\
	\hline 
		\hline
		\end{tabular*}
	\caption{ (Central column) Reflection coefficient at the bottom of the conduction band. (Right column) SF energy resonances from the bottom of the conduction band at the $M$ Brillouin-zone point.}
	\label{tab:reflect}
\end{table}

Results regarding cubic SiC show that the physical extend of the stacking fault can be extremely important for its scattering properties, giving rise to electronic reflections that range from partial to total. Qualitatively, the same mechanism lies underneath the coherent scattering process in hexagonal SiC polytypes. As an example, here we simulate the extended (4,4) Shockley-type defect \cite{2011APExp...4b5802C} [Fig. \ref{fig:4hcond}(c)] that appears to be one of the most common SFs in 4H-SiC crystals \cite{2008ApPhL..92v1906F}. The methodology remains identical as in the case of 3C-SiC, considering now that the SF is perpendicular to the [0001] direction. The computational outcome [Fig. \ref{fig:4hcond}(b)] indicates a severe transport damage for the conduction band and a respective increase of the material's resistivity along the c-axis for $n$-doped systems. Such feature is in excellent agreement with the experiment \cite{1997JCrGr.181..229T,2003PhR...379..149F}, indicating that resonant scattering effects are indeed at the origin of the transport anisotropy in SiC crystals. Similar phenomena have been theoretically predicted for III-V nitride semiconductors within a semiclassical picture \cite{2011ApPhL..98b2109K}.

A final issue that needs discussion is the weight of coherent  backscattering with respect to phase incoherency and other diffusive scattering mechanisms that exist in the electrical conduction of realistic 3C-SiC crystals. We can expect that dephasing of the electronic waves should flatten the conductance distribution and smear its  maxima and minima. Moreover, dissipative events (e.g. electron-phonon interactions) are expected to further increase the resistance with respect to the ideal case, tuning the electrical conduction according to Ohm's law.  However, the impact of the coherent backscattering mechanism for energies around the minimum of the conduction band is expected to be prevailing for extended SFs. The angle of the incident electron waves with respect to the SF is again an important factor for this phenomenon. The results shown here ($\phi=90^{\circ}$) represent a condition of maximum backscattering. For $90^{\circ} < \phi < 0^{\circ}$ gradual leakage currents are expected to pass through the SF barriers and increase the conductance. In the case of $\phi=0^{\circ}$, .i.e. when SFs are parallel to the transport direction, we can expect an absence of any resonant scattering effect. In this case the increased DOS of SFs should give rise to a higher localized current density with respect to that of the rest of the crystal (in the case when the SF is contacted by a metallic electrode). Such characteristic could explain recent current maps obtained by $n$-type 3C-SiC structures using conductive atomic force microscopy \cite{2009ApPhL..95h1907E}. 

To conclude, in this letter we have investigated the electronic conduction properties of SFs in 3C- and 4H-SiC within first-principles quantum transport calculations. We have found that SFs strongly scatter the electron waves at particular energies and Brillouin zone subspaces for both the conduction and the valence band. However, a significant increase in the material's resistivity is only expected for $n$-type systems and incident electron angles other than $\phi=0^{\circ}$.  In addition, the presence of extended SFs should totally reflect the electron waves at the bottom of the conduction band, effectively increasing the height of the measured Schottky-barrier. Based on these results we argue that, contrary to silicon, electron resonant scattering from SFs should be one of the principal sources of instability and device degradation in silicon carbide.

I.D. and A.L. acknowledge the European Science Foundation (ESF) under the EUROCORES Program EuroGRAPHENE CRP GRAPHIC-RF for partial financial support, as well as the CINECA consortium for computational resources (project SUBGRAPH).

\bibliography{sic-sf}


\end{document}